# The Benefits of Data Storytelling in Accessible Teaching

A conceptual framework to meet the requirements of Title II of the ADA Act

Marina Buzzi
IIT-CNR
Pisa, Italy
marina.buzzi@iit.cnr.it

Barbara Leporini
University of Pisa
ISTI-CNR
Pisa, Italy
barbara.leporini@unipi.it

Angelica Lo Duca
IIT-CNR
Pisa, Italy
angelica.loduca@iit.cnr.it

## ABSTRACT

Accessible teaching has been extensively investigated in computer science, yet its integration into other disciplines, such as data literacy, remains limited. This paper examines the potential of data storytelling, defined as the integration of data, visualizations, and narrative, as a possible strategy for making complex information accessible to diverse learners in compliance with Title II of the Americans with Disabilities Act (ADA). We propose six design principles, derived from Title II's core obligations, to guide educators in applying data storytelling within inclusive learning environments. A simulated scenario shows the operationalization of these principles, illustrating how narrative-driven data presentation can enhance comprehension, engagement, and equitable access across different educational contexts.

## 1 Introduction

Accessible teaching aims to design environments, resources, and strategies that allow all students to fully engage with and benefit from instruction, regardless of physical, sensory, cognitive, or learning differences [1-4]. Inclusive teaching frameworks such as Universal Design for Learning (UDL) promote multiple means of representation, engagement, and expression to ensure that content is perceivable and usable through various modalities. In parallel, updated guidance under Title II of the Americans with Disabilities Act (ADA) makes clear that digital learning resources must be accessible to people with disabilities (U.S. Department of Justice, 2022).

Beside accessibility, learner engagement has emerged as a critical factor in effective education. Research highlights that students are more likely to achieve deep and sustained learning when instructional design incorporates active participation, emotional investment, and cognitive challenge [5]. Accessible teaching, when combined with strategies that foster engagement, creates environments in which learning is personally meaningful and intrinsically motivating.

While accessibility is widely recognized in disciplines such as computer science, its systematic application to other areas, such as data literacy, remains underexplored [6]. Data literacy, the ability to interpret, evaluate, and communicate data effectively, is increasingly vital in academic, professional, and civic contexts. Yet, the abstract and often technical nature of data can create significant comprehension barriers, particularly for learners with disabilities or those from non-technical backgrounds. This paper proposes data storytelling as a powerful means of addressing these challenges. More specifically, this paper focuses on how data storytelling can be made accessible for students with special needs. Data storytelling embeds data within a coherent narrative arc and adapts its delivery to multiple modalities [7]. To demonstrate conceptually how data storytelling can be used for accessible teaching, we extract six design principles from the requirements of Title II of the Americans with Disabilities Act (ADA) and describe how they can be applied to data storytelling. The six design principles include equitable access, effective communication, programmatic accessibility, integration, reasonable modifications, and auxiliary aids and services. We also propose a simulated scenario where the proposed principles are applied.

The paper is organized as follows. In Section 2, we review existing research on accessible teaching and data storytelling. In Section 3, we present six design principles for aligning storytelling with ADA Title II obligations and illustrate their application through a simulated teaching scenario (Section 4). Finally, Section 5 describes conclusions and future work.

## 2    Related Work

Data storytelling integrates data, visual representations, and narrative techniques to convey insights in a compelling and accessible way [8].

In education, this approach has gained traction over the past decade as a means to enhance data literacy and learner engagement across all levels. Data storytelling moves beyond data visualization by embedding quantitative evidence within a narrative arc, using elements such as context, characters, and plot to make information more relatable and comprehensible [9].

A growing body of research positions data storytelling as both an instructional strategy and an authentic assessment tool. McDowell and Turk describe a graduate information science course where students iteratively created data stories on social issues, shifting from consumers to producers of data interpretations and developing advanced analytical and critical thinking skills [10]. At the undergraduate level, Li et al. implemented the "OCEL.AI" model, an open collaborative experiential learning approach built around storytelling, which improved both data literacy and career motivation in data science [11]. In K–12 contexts, Stornaiuolo (2020) reported on a high school makerspace where students transformed personal or community datasets into visual narratives, such as infographics printed on T-shirts, fostering creativity alongside analytic reasoning [12].

Empirical studies have also examined cognitive and motivational impacts of data storytelling in education. Shao et al. found that adding narrative elements to visualizations significantly improved participants' efficiency and accuracy in comprehension tasks, particularly for extracting key insights [13].

The broader literature links these pedagogical benefits to accessibility and inclusivity goals. Data storytelling's multimodal nature, which combines visual, textual, auditory, and sometimes interactive elements, aligns with UDL principles and ADA Title II requirements by offering multiple pathways to engagement. In a previous study, we have already investigated the potential benefits of data storytelling for learners with special needs [14]. That work proposed a methodology integrating UDL and AI-assisted storytelling to adapt general education programs into personalized, multisensory narratives tailored to each student's needs and available assistive technologies. The present study aims to generalize the preliminary results from a broader and more theoretical perspective.

## 3    Design Principles for Accessible Data Storytelling

The design principles we propose in this paper are specifically designed to meet the requirements of Title II of the Americans with Disabilities Act (ADA). More specifically, we extracted six strict requirements from 28 CFR Part 35 of Title II and declined them for educational contexts. The six core obligations include equitable access, effective communication, programmatic accessibility, integration, reasonable modifications, and auxiliary aids and services.

**Equitable Access**: All learners, regardless of disability, must have access to instructional content and learning environments.
**Effective Communication**: Information must be conveyed in ways that are understandable and usable, with necessary aids and services provided.
**Programmatic Accessibility**: Educational offerings must be accessible as a whole, not only through separate or alternate channels.
**Integration**: Students with disabilities must be able to participate alongside their peers in mainstream environments.
**Reasonable Modifications**: Adjustments to teaching methods or materials must be made when necessary to ensure access.
**Auxiliary Aids and Services**: Institutions must provide assistive technologies or services (e.g., captions, screen readers, interpreters) where needed.

These principles establish a legal and ethical baseline for inclusive education. However, making them operative, outside of disability-focused disciplines, remains a significant pedagogical challenge. To address this gap, we propose a set of design strategies that leverage data storytelling to fulfill the requirements of Title II. This paper does not focus on teaching accessibility (i.e., instructing students about accessible design). We reserve the analysis of this aspect to future work.

Data storytelling involves the combination of narrative, data, and visual elements to lead and support understanding. Unlike raw data presentations or static dashboards, narrative formats provide structure and cognitive support and can be modularly adapted to different formats and user needs.

Table 1 maps each Title II requirement to a corresponding design strategy in accessible data storytelling.

**Table 1. Mapping ADA Title II requirements to data storytelling design practices**

| ADA Title II Principle | Design Strategy |
| --- | --- |
| Equitable Access | Content is adaptable across multiple modalities. |
| Effective Communication | Data is contextualized for improved comprehension. |
| Programmatic Accessibility | Narrative can be delivered via accessible digital channels. |
| Integration | Interaction pathways support inclusive participation. |
| Reasonable Modifications | Narrative depth and pacing are tunable to learner needs. |
| Auxiliary Aids and Services | Co-design ensures compatibility with different assistive technologies. |

Equitable access to story content can be delivered in multiple formats (e.g., audio, text, images, animation), ensuring access for diverse learners. Narrative content is modular and can be adapted post-design to fit sensory, cognitive, or technological constraints, based on the principle that the creation process of a story is separated from the delivery phase [8].

Storytelling can reduce cognitive barriers by contextualizing meaning. Anchoring abstract data in a structured story can improve clarity and retention, particularly for learners with limited numeracy or unfamiliarity with the topic (effective communication).

The storytelling layer is format-agnostic: instructors can select the medium of delivery (e.g., tactile, auditory, or interactive) most suited to the student's context while preserving the same core narrative logic (Programmatic Accessibility). In addition, storytelling interfaces can include interactive components that allow all students to engage with the same narrative structure, navigating content through personalized, discipline-specific paths without exclusion (Integration). Story arcs can also be adjusted in granularity or complexity to meet diverse cognitive needs, for example, simplifying exposition, adjusting pacing, or offering scaffolding for foundational concepts (Reasonable Modifications). Finally, stories should be co-designed with users who rely on assistive technologies. Incorporating feedback from disabled learners ensures narrative coherence across modalities such as screen readers, haptics, or voice interfaces (Auxiliary Aids and Services).

The proposed framework positions data storytelling not merely as a communication technique but as a powerful accessibility strategy. By separating the design of the narrative (semantic structure) from its delivery (technical rendering) [8], educators can meet the functional diversity of learners without compromising narrative coherence. Furthermore, these principles align with Universal Design for Learning (UDL) guidelines, offering a scalable approach for accessibility across disciplines.

## 4 Simulated Scenario

To demonstrate how the proposed design principles potentially work practically, we implemented a simulated scenario. At the moment, we have not had them implemented into a real case study. However, the simulated scenario can serve as a proof of concept of the proposed design principles.

In the following, first we describe the simulated scenario, and then we discuss how it implements the proposed design principles.

### 4.1 Description

To help students understand global earthquake activity (magnitude > 4.5) from 2023 to 2024, a teacher constructs a simulated lesson using a three-act data storytelling structure. The activity is based on real data visualized in Figure 1, sourced from the USGS (U.S. Geological Survey) Earthquake Catalog [15].

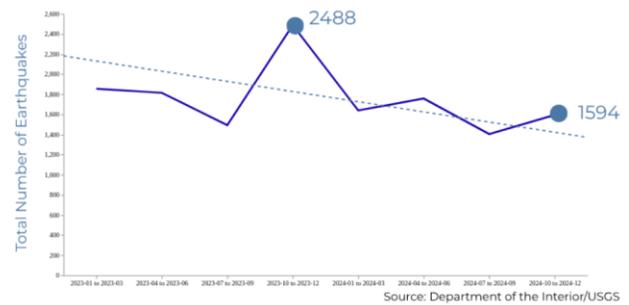

**Figure 1: The trendline related to the global earthquake activity (magnitude > 4.5) from 2023 to 2024.**

The teacher organizes the lesson using the three-act structure described below.

**First Act: The Setup.** Students are introduced to the concept of seismic activity and the importance of monitoring earthquake patterns. The teacher poses an open-ended question: "Have earthquakes been increasing or decreasing lately? What might that mean?" Students are shown the visual (Figure 1), but without any annotations. They are asked to describe what they notice.

For accessibility, the visual is presented in multiple formats, including a tactile graph for blind learners, an audio narration describing the chart's ups and downs, and a simplified table with interval data [16].

**Second Act: The Conflict.** The teacher highlights a surprising peak in the graph: 2,488 earthquakes between October and December 2023. Students break into groups (with accessibility support such as tactile graph or audio narration) and investigate potential global events, such as volcanic eruptions, tectonic shifts, or data anomalies. The narrative prompts curiosity, making the abstract numbers meaningful.

**Third Act: The Resolution.** Students return to the original chart, now annotated with a regression line showing a slight downward trend overall, ending at 1,594 events in the last interval. The teacher guides a discussion by posing questions to the students, such as the following ones: Is the spike an outlier? How should we interpret variability in natural systems? Students propose explanations and policy implications (e.g., emergency preparedness, climate effects). Optional scaffolding is offered: pacing aids for learners with processing delays, or simplified summaries for neurodiverse students.

### 4.2 Discussion

The simulated earthquake lesson illustrates how data storytelling can embody the six core obligations of ADA Title II, transforming compliance requirements into inclusive pedagogical design.

**Equitable Access.** The earthquake narrative is accessible through multiple modalities, including visual, auditory, tactile, and textual. The activity supports learners with sensory, cognitive, or technological constraints. For example, the chart is accompanied by an audio description and a simplified data table, ensuring no single format is a barrier.

**Effective Communication.** Rather than presenting raw data, the lesson anchors numerical values within a narrative arc. This structure helps students, especially those with limited numeracy, grasp abstract concepts like variability and trend analysis. Posing guiding questions and contextualizing spikes (e.g., the 2,488 earthquakes) enhances comprehension and reduces cognitive load.

**Programmatic Accessibility.** The narrative framework is delivery-neutral: it can be deployed in person, remotely, or through assistive platforms. Whether students use a screen reader, a Braille display, or a standard laptop, the core storyline remains intact, and content becomes understandable. This ensures that accessibility is not limited to "special versions" of the lesson but is baked into the instructional design.

**Integration.** All students participate in the same activity, regardless of ability. Group discussions, guided inquiries, and visual annotations are designed to be inclusive, avoiding separate or simplified tracks for disabled students. For example, haptic tools and visual organizers support different access needs without fragmenting the learning experience.

**Reasonable Modifications.** The teacher adapts the story depth and pacing to match the learner's needs. Complex patterns in the data can be explored gradually. Students with attention or processing differences benefit from modular, step-by-step exploration of the narrative.

**Auxiliary Aids and Services.** The lesson design anticipates assistive technology use from the outset. Narrations are compatible with screen readers; visuals are described using alt-text; and tactile versions of the chart are provided.

## 5    Conclusions and Future Work

The proposed approach illustrates how data storytelling is not only a pedagogical enhancement but also a concrete method for fulfilling legal accessibility requirements. By decoupling content structure from delivery format, educators can create inclusive learning environments where all students can access, interpret, and engage with complex data on equal footing, such as global seismic activity.

However, our findings are currently based on a simulated teaching scenario. As future work, we plan to implement and evaluate the proposed framework in real classroom environments across diverse educational contexts and disciplines. This will allow us to assess its practical feasibility, measure its impact on learner outcomes, and refine the design principles based on empirical evidence and user feedback. We are also investigating how AI-assisted narrative generation can further expand the reach and effectiveness of accessible data storytelling [14].

## ACKNOWLEDGMENTS

This study is part of the PRIN project 2022HXLH47 "STEMMA-Science, Technology, Engineering, Mathematics, Motivation and Accessibility" (funded by the European Union- Next Generation EU, Mission 4 Component C2 CUP B53D23019500006).

## REFERENCES

1. Yang, C. (2025). Adapting teaching methods to accommodate diverse learning styles in education. Journal of Higher Education Research. https://doi.org/10.32629/jher.v5i6.3382.
2. Debasu, H., & Yitayew, A. (2024). Examining elements of designing and managing of creating inclusive learning environment: Systematic literature review. *International Journal of Special Education*, *39*(1), 33-43.
3. Dziatkovskaya, E. N., Dlimbetova, G. K., & Dziatkovskii, A. D. (2021). Accessible education: designing the educational environment for sustainable development.
4. Hitchcock, C. (2001). Balanced instructional support and challenge in universally designed learning environments. Journal of Special Education Technology, 16(4), 23-30.
5. Capone, R. (2022). Blended learning and student-centered active learning environment: A case study with STEM undergraduate students. *Canadian Journal of Science, Mathematics and Technology Education*, *22*(1), 210-236.
6. Marriott, K., Lee, B., Butler, M., Cutrell, E., Ellis, K., Goncu, C., Hearst, M., McCoy, K., & Szafir, D. A. (2021). Inclusive data visualization for people with disabilities: A call to action. Interactions, 28(3), 47–51. https://doi.org/10.1145/3457875
7. Dykes, B. (2019). Effective data storytelling: how to drive change with data, narrative and visuals. John Wiley & Sons.
8. Lo Duca, A (2025). Become a Great Data Storyteller. John Wiley and Sons.
9. Lo Duca, A., & McDowell, K. (2025). Using the S-DIKW framework to transform data visualization into data storytelling. Journal of the Association for Information Science and Technology, 76(5), 803-818.
10. McDowell, K., & Turk, M. J. (2024). Teaching data storytelling as data literacy. Information and Learning Sciences, 125(5/6), 321-345.
11. Li, Y., Wang, Y., Lee, Y., Chen, H., Petri, A. N., & Cha, Y. (2023). Teaching data science through storytelling: Improving undergraduate data literacy. Thinking Skills and Creativity, 48, 101311.
12. Stornaiuolo, A. (2019). Authoring Data Stories in a Media Makerspace: Adolescents Developing Critical Data Literacies. Journal of the Learning Sciences, 29(1), 81–103. https://doi.org/10.1080/10508406.2019.1689365
13. Shao, H., Martinez-Maldonado, R., Echeverria, V., Yan, L., & Gasevic, D. (2024). Data storytelling in data visualisation: Does it enhance the efficiency and effectiveness of information retrieval and insights comprehension? In Proceedings of the 2024 CHI Conference on Human Factors in Computing Systems (Article 195, pp. 1–21). https://doi.org/10.1145/3613904.3643022
14. Buzzi, M., Leporini, B., Lo Duca, A., Punzo, V., Rotelli, D. (2025). Inclusive Data Literacy: UDL and AI-Assisted Data Storytelling for BVI Students. In: Cristea, A.I., Walker, E., Lu, Y., Santos, O.C., Isotani, S. (eds) Artificial Intelligence in Education. Posters and Late Breaking Results, Workshops and Tutorials, Industry and Innovation Tracks, Practitioners, Doctoral Consortium, Blue Sky, and WideAIED. AIED 2025. Communications in Computer and Information Science, vol 2592. Springer, Cham. https://doi.org/10.1007/978-3-031-99267-4_6
15. U.S. Geological Survey. Earthquake Catalog. Retrieved from: https://earthquake.usgs.gov/earthquakes/search/ (2025-07-17)
16. Diagram Center. Retrieved from http://diagramcenter.org/making-images-accessible.html (2025-08-13)